\documentclass[12pt]{article}
\usepackage{amsmath,amssymb}
\usepackage{graphicx,color}
\numberwithin{equation}{section}
\usepackage{cite}
\usepackage{bm}
\usepackage{dcolumn}
\newcommand{\be}{\begin{equation}}
\newcommand{\bea}{\begin{eqnarray}}
\newcommand{\eea}{\end{eqnarray}}
\newcommand{\ba}{\begin{array}}
\newcommand{\ea}{\end{array}}
\newcommand{\ee}{\end{equation}}

\expandafter\ifx\csname mathbbm\endcsname\relax

\else

\fi \textheight 22cm \textwidth 15cm \topmargin 1mm \oddsidemargin
5mm \evensidemargin 5mm

\begin{document}
\begin{titlepage}
\hfill \vbox{
    \halign{#\hfil         \cr
         \cr
                      } % end of \halign
      }  % end of \vbox
\vspace*{20mm}
\begin{center}
{\Large {\bf The minimum energy state }\\
}

\vspace*{15mm} \vspace*{1mm} {Amin Akhavan}

\vspace*{.4cm}

{\it  School of Particles and Accelerators, Institute for Research in Fundamental Sciences (IPM)\\
P.O. Box 19395-5531, Tehran, Iran\\$email:amin_- akhavan@ipm.ir$ }

\vspace*{2cm}

\end{center}

\begin{abstract}
We define the minimum energy state while the expectation value of the field, evolves in time. We obtain the relation between the n-point functions in such a state, and the external field for all the moments. We obtain an equation of motion and the renormalization counterterms for the external field in the first order of interaction.     

\end{abstract}

\vspace{2cm}

\end{titlepage}

\section{Introduction}
One of the interesting interpretation of the effective action has been introduced by K.Symanzki, is the minimum of the Hamiltonian expectation value while the fields expectation values are constrained\cite{zymanski}. In this interpretation, the vacuum state will relax to a state in which the effective action is not only stationary but also minimum. An important consequence is that the effective potential curve in the vacuum, has to be convex. The symmetry breaking method has obtained from this viewpoint.\\ In an example of the actions with $m^2<0$, the effective action has two minimum points, but a negative-definite second derivative between them. But if the expectation value of the field is a value between minima -in the non zero source- the effective action has a concave curve in which has a contradiction to minimum energy interpretation. To solve this, we have to consider the effective potential as a constant value between the two minima, satisfying the effective potential to be convex\cite{weinberg}.\\
In continue one can define an axiom that all the ground states are in the minimum energy constrained to the background fields, before they get excited.\\
In the Symanzik method, the Lagrange multipliers mechanism are used such that the non dynamic multipliers related to the sources turned on in the action. In this method: the expectation values of a field operator applied in the equations of the turned on sources, have to be equal to the expectation value of the field operator applied in the equations with the turned off sources. Therefore the expectation values are not dynamic:\\
\be
<\Omega_{J}|\Phi(t,\vec{x})|\Omega_{J}>=<\Omega_{J}|\Phi_{J}(t,\vec{x})|\Omega_{J}>=\phi_{J}(\vec{x})
\ee           
In fact, if the expectation values of the field would be dynamic, the Lagrange multipliers could not be related completely to the sources turned on in the generator function $W(J)$, and the effective action could not be proportional to the hamiltonian expectation value.\\
Another content we have focused on, is the macroscopic external fields measured in a classical way. If an external field is not dynamic, we can consider it as the expectation value of the field operator in the vacuum state:\\
\be
\phi(\vec{x})=<\Omega|\Phi(t,\vec{x})|\Omega>
\ee
And if the Hamiltonian and the momentum operators are reversible, the external field is constant. In this consideration, the equation of the variation of the effective action apply on the external field, but such a trivial application, because such the spacial solution is not dynamic. What about the general solutions?\\
It seems that the real dynamic external fields have to be defined from the quantum states that they live in, but the variation of the effective actions determine only the sources (in the generator function). For example the variation of the action, shows that the source of the real vacuum expectation value is zero, and not that, in what equation of motion the vacuum expectation value of the field applied. And for this reason, we have to consider the effective actions, just as the generator functions.\\
But now we can ask, in what quantum state, the measurement of the dynamical external fields, occurred.\\
In this paper, in the second section we define the axiom of minimum energy for the dynamic external fields. In the third section we solve an example and obtain the equation of motion for the external field untill the first order of interaction. In the last section, we have a discussion.  
\section{The minimum energy state}
We want to define a state $|\psi_{0}>$, in which the fields can be measured in classical way:
\be
\phi(t,\vec{x})=<\psi_{0}|\Phi(t,\vec{x})|\psi_{0}>.
\ee
For dynamic value we can not consider $|\psi_{0}>$, as a hamiltonian eigen state. Now we consider the axiom of the minimum energy, we have offered in the introduction. For this purpose, we have to use constraints for the external field. We do not want to determine the field in all times, we only want to locate the external field on the determined time evolution path. For this job we need to constrain the field and its time derivation in a fixed time $t_{0}$: 
\be
\phi(t_{0},\vec{x})=\varphi_{0}(\vec{x})~~~~~\dot{\phi}(t_{0},\vec{x})=\dot{\varphi}_{0}(\vec{x}).
\ee
Including these constraints and$<\psi_{0}|\psi_{0}>=1$ and using the Lagrange multipliers method, we minimize the expectation value of the Hamiltonian:
\bea
\delta <\psi_{0}|H|\psi_{0}>-\int d^3xJ_{0}(\vec{x}) \delta<\psi_{0}|\Phi(t_{0},\vec{x})|\psi_{0}>-\int d^3xK_{0}(\vec{x})<\psi_{0}|\dot{\Phi}(t_{0},\vec{x})|\psi_{0}>\nonumber\\
-E\delta<\psi_{0}|\psi_{0}>=0~~~~~~~~~~~~~~~~~~~~~~~~~~~~
\eea
therefore:
\be
\int d^3x\bigg(\mathcal{H}[\Phi(t_{0},\vec{x}),\dot{\Phi}(t_{0},\vec{x})]-J_{0}(\vec{x})\Phi(t_{0},\vec{x})-K_{0}(\vec{x})\dot{\Phi}(t_{0},\vec{x})\bigg)|\psi_{0}>=E|\psi_{0}>.
\ee
Now, we can see that, $|\psi_{0}>$ is the minimum eigenstate of the operator has been written. We show this operator, $H^{J,K}$ defined at the time $t_{0}$. \\
For the models in which the functional of $\dot{\Phi}(t_{0},\vec{x})$ in their Hamiltonian is quadratic, we can write the new operator like this:
\be
H^{J,K}=\int d^3x\mathcal{H}^{J,K}=\int d^3x \frac{1}{2}\pi^2(\vec{x})+V[\Phi(t_{0},\vec{x})]-J_{0}\Phi(t_{0},\vec{x})
\ee 
while, 
\be
\pi(\vec{x})=\dot{\Phi}(t_{0},\vec{x})-K_{0}(\vec{x})
\ee
and
\be
[\Phi(t_{0},\vec{x}),~\pi(\vec{x'})]=i\delta^3(\vec{x}-\vec{x'}).
\ee
According to the appearance of $H^{J,K}$, we conclude that  
\be
<\psi_{0}|\pi|\psi_{0}>=0
\ee
and therefore :~~ $K_{0}(\vec{x})=\dot{\varphi}_{0}(\vec{x})$.\\
In continue,$|\psi_{0}>$ and $J_{0}(\vec{x})$ are the other unknowns of our problem. Since we want to obtain the external field in the equation(2.1), we need to find \\ \be\phi(t,\vec{x})=<\psi_{0}|\Phi(t,\vec{x})|\psi_{0}>=<\psi_{0}|e^{i(t-t_{0})H}\Phi(t_{0},\vec{x})e^{-i(t-t_{0})H}|\psi_{0}>
\ee
as the n-point functions of the operators at the moment $t_{0}$, in the minimum energy state. 
To obtain these types of n-point functions through the Feynman diagrams - provided we do not run out of the time $t_{0}$- we have to consider the state $|\psi_{0}>$, like this:
\be 
\lim_{S \to \infty (1+i\epsilon)} e^{iSH^{J,K}}|0>=\lim_{S \to \infty (1+i\epsilon)} e^{iSE}|\psi_{0}>
\ee
in which $S$ is just a parameter. Also we define a Heisenberg picture along the parameter $s$ , like this:
\be 
\Phi_{H}(s,\vec{x})=e^{isH^{J,K}}\Phi(t_{0},\vec{x})e^{-isH^{J,K}}\nonumber\\
\ee
\be
\pi_{H}(s,\vec{x})=e^{isH^{J,K}}\pi(\vec{x})e^{-isH^{J,K}}=\frac{d\Phi_{H}(s,\vec{x})}{ds}.
\ee
Here, the state $|0>$ is the vacuum state of the free Hamiltonian:
\be
\int d^3x ~~\frac{1}{2}\pi^2_{I}(s,\vec{x})+ \frac{1}{2}m^2 \Phi^2_{I}(s,\vec{x})
\ee
where $\pi_{I}$ and $\Phi_{I}$ are the operators in the interaction picture along the\\ parameter $s$.\\
In this case, the all n-point functions can be written like this:
\be
<\psi_{0}|A[\Phi(t_{0},\vec{x})]|\psi_{0}>=<\psi_{0}|e^{isH^{J,K}}A[\Phi(t_{0},\vec{x})]e^{-isH^{J,K}}|\psi_{0}>\nonumber\\
\ee
\be
=<\psi_{0}|A[\Phi_{H}(s,\vec{x})]|\psi_{0}>=\frac{<0|A[\Phi_{I}(s,\vec{x})]e^{-i\int dsd^3x\mathcal{H}_{I}^{J,K}}|0>}{<0|e^{-i\int dsd^3x\mathcal{H}_{I}^{J,K}}|0>}
\ee
since  
\be
<\psi_{0}|\Phi_{H}(s,\vec{x})]|\psi_{0}>=<\psi_{0}|\Phi(t_{0},\vec{x})]|\psi_{0}>=\varphi_{0}(\vec{x}).
\ee
Therefore we define
\be
<\psi_{0}|H^{J,K}[\Phi_{H}(s,\vec{x})]|\psi_{0}>=\frac{\Gamma[\varphi_{0}(\vec{x})]}{S}.
\ee
This is a definition of effective action for the external fields which their coordinates are only spacial\cite{zymanski,weinberg}, and trivially we can find that:
\be
J_{0}=-\frac{\delta \Gamma(\varphi_{0})}{\delta \varphi_{0}}.
\ee
Now, Using the equations (2.9) and (2.13), $\phi(t,\vec{x})$ could be obtain.\\
In continue, we want to find the equation of motion of $\phi(t,\vec{x})$.\\ The equation of motion for the operator $\Phi(t,\vec{x})$ is:
\be
\ddot{\Phi}(t,\vec{x})-\nabla^2\Phi(t,\vec{x})+\frac{\delta V[\Phi(t,\vec{x})]}{\delta \Phi(t,\vec{x})}=0.
\ee
Considering the expectation value of the equation and being at the desired moment $t_{0}$, we have:
\be
\ddot{\phi}(t_{0},\vec{x})-\nabla^2\phi(t_{0},\vec{x})+<\psi_{0}|\frac{\delta V[\phi(t_{0},\vec{x})]}{\delta\phi(t_{0},\vec{x})}|\psi_{0}>=0.
\ee
If we solve the n-point function $<\psi_{0}|\frac{\delta V[\phi]}{\delta\phi}|\psi_{0}>$; finally we obtain the equation of motion of the external field, but at the moment $t_{0}$.\\
Assume that we do all the computations have been done again at the moment $t_{1}=t_{0}+\epsilon$, therefore the equation (2.18) applies on $\phi_{1}(t_{1},\vec{x})$ and $|\psi_{1}>$ with these new constraints:
\be
\phi_{1}(t_{1},\vec{x})= <\psi_{1}|\Phi(t_{1},\vec{x})|\psi_{1}>=\varphi_{1}(\vec{x})\nonumber\\ 
\ee
\be
\dot{\phi_{1}}(t_{1},\vec{x})= <\psi_{1}|\dot{\Phi}(t_{1},\vec{x})|\psi_{1}>=\dot{\varphi}_{1}(\vec{x})
\ee
and $|\psi_{1}>$  would be the minimum energy state for the new constraints. But if we want the new constraints to be the same as the previous ones, and therefore $\phi_{0}(t_{0})$ and $\phi_{1}(t_{1})$ are placed on the time line of a determined solution, we have to chose  $\varphi_{1}(\vec{x})$ and $\dot{\varphi}_{1}(\vec{x})$ in the equations (2.19), such that:
\be
\phi_{1}(t_{1},\vec{x})=\phi(t_{0},\vec{x})+\epsilon\dot{\phi}_{0}(t_{0},\vec{x})\nonumber\\
\ee
\be
\dot{\phi}_{1}(t_{1},\vec{x})=\dot{\phi}(t_{0},\vec{x})+\epsilon\ddot{\phi}_{0}(t_{0},\vec{x})
\ee
therefore:
\be
<\psi_{1}|\Phi(t_{1}, \vec{x})|\psi_{1}>=<\psi_{0}|\Phi(t_{1}, \vec{x})|\psi_{0}>\nonumber\\
\ee
\be
<\psi_{1}|\dot{\Phi}(t_{1}, \vec{x})|\psi_{1}>=<\psi_{0}|\dot{\Phi}(t_{1}, \vec{x})|\psi_{0}>.
\ee
If the equation (2.18) finally is a time local quadratic differential equation, we can expand the equations (2.20) and (2.21) to all time derivatives, and therefore:
\be
|\psi_{0}>=|\psi_{1}> ~and ~~ \phi_{1}(t_{1},\vec{x})= \phi(t_{1},\vec{x}).
\ee
In this case, the equation of motion (2.18) at the time $t_{0}$, will be corrected for all times.
\section{Scalar field example}
Here for a scalar field with an interaction term, $\frac{\lambda}{4!}\Phi^4$, we will obtain the equation of motion.\\
Considering the Lagrangian density like this:
\bea
\mathcal{L}=\frac{1}{2}\partial_{\mu}\partial^{\mu}\Phi-\frac{1}{2}m^2\Phi^2-\frac{\lambda}{4!}\Phi^4\nonumber\\
+\frac{1}{2}\delta_{z}\partial_{\mu}\partial^{\mu}\Phi-\frac{1}{2}\delta_{m}\Phi^2-\frac{\delta_{\lambda}}{4!}\Phi^4
\eea
and using the equation(2.5), we have:
\bea
\mathcal{H}^{J,K}=\frac{1}{2}\dot{\Phi}^2(t_{0},\vec{x})+\frac{1}{2}(\vec{\nabla}\Phi(t_{0},\vec{x}))^2+\frac{1}{2}m^2\Phi^2(t_{0},\vec{x})+\frac{\lambda}{4!}\Phi^4(t_{0},\vec{x})\nonumber\\
-\frac{1}{2}\delta_{z}\partial_{\mu}\Phi(t_{0},\vec{x})\partial^{\mu}\Phi(t_{0},\vec{x})+\frac{1}{2}\delta_{m}\Phi^2(t_{0},\vec{x})+\frac{\delta_{\lambda}}{4!}\Phi^4(t_{0},\vec{x})\nonumber\\
-J_{0}(\vec{x})\Phi(t_{0},\vec{x})-K_{0}(\vec{x})\dot{\Phi}(t_{0},\vec{x})-\delta J_{0}(\vec{x})\Phi(t_{0},\vec{x})-\delta K_{0}(\vec{x})\dot{\Phi}(t_{0},\vec{x}).
\eea
We will separate  the field operators:
\be
\Phi(t_{0},\vec{x})=\varphi_{0}(\vec{x})+\chi(\vec{x})\nonumber\\
\ee
\be
\dot{\Phi}(t_{0},\vec{x})=\dot{\varphi}_{0}(\vec{x})+\pi(\vec{x})
\ee
in which the constraints will be written like this:
\be
<\psi_{0}|\chi(\vec{x})|\psi_{0}>=0 ~~and ~~<\psi_{0}|\pi(\vec{x})|\psi_{0}>=0.
\ee 
Using the second constraint, we will have:
\bea
\mathcal{H}^{J,K}=\frac{1}{2}\pi^2(\vec{x})+\frac{1}{2}(\nabla\chi(\vec{x}))^2+\frac{1}{2}m^2\chi^2(\vec{x})+\frac{\lambda}{4!}(\varphi_{0}(\vec{x})+\chi(\vec{x}))^4\nonumber\\
-\frac{1}{2}\delta_{z}\pi^2(\vec{x})+\frac{1}{2}\delta_{z}(\nabla\chi(\vec{x}))^2+\delta_{z}\nabla\chi(\vec{x}).\nabla\varphi_{0}(\vec{x})\nonumber\\
+\frac{1}{2}\delta_{m}(\varphi_(\vec{x})+\chi(\vec{x}))^2+\frac{\delta_{\lambda}}{4!}(\varphi_{0}(\vec{x})+\chi(\vec{x}))^4-\delta J_{0}(\vec{x})\chi(\vec{x})+const
\eea
and also
\be
K_{0}(\vec{x})=\dot{\varphi}_{0}(\vec{x}) ~~and ~~\delta K_{0}(\vec{x})=-\delta_{z}\dot{\varphi}_{0}(\vec{x}) ~~and ~~considering~~J_{0}(\vec{x})=-\nabla^2\varphi_{0}(\vec{x})+m^2\varphi_{0}(\vec{x}).
\ee
Also we can write it in the interaction picture along the parameter $s$:
\bea
\mathcal{H}^{J,K}_{I}(s,\vec{x})=\frac{\lambda}{4!}(\varphi_{0}(\vec{x})+\chi_{I}(s,\vec{x}))^4
-\frac{1}{2}\delta_{z}(\frac{d\chi_{I}(s,\vec{x})}{ds})^2 \nonumber\\+\frac{1}{2}\delta_{z}(\nabla\chi_{I}(s,\vec{x}))^2+\delta_{z}\nabla\chi_{I}(s,\vec{x}).\nabla\varphi_{0}(\vec{x})\nonumber\\
+\frac{1}{2}\delta_{m}(\varphi_{0}(\vec{x})+\chi_{I}(s,\vec{x}))^2+\frac{\delta_{\lambda}}{4!}(\varphi_{0}(\vec{x})+\chi_{I}(s,\vec{x}))^4\nonumber\\-\delta J_{0}(\vec{x})\chi_{I}(s,\vec{x})+const.
\eea
To find~ $\delta J_{0}(\vec{x})$, we need to use the first constraint:
\be
<\psi_{0}|\chi(\vec{x})|\psi_{0}>=<\psi_{0}|\chi_{H}(s,\vec{x})|\psi_{0}>=0.
\ee
(Where $\chi_{H}(s,\vec{x})$, is the Heisenberg picture along the parameter $s$.)
To do so, we have to use the equation (2.13) and (3.8), but we want to obtain it until the first order of the interaction coefficient $\lambda$.
Therefore, we consider the odd power terms of $\chi$ in $\mathcal{H}^{J,K}$ to write the first order of expectation value:
\bea
<\psi_{0}|\chi_{H}(s,\vec{x})|\psi_{0}>^{(\lambda)}=\int ds'd^3x'\frac{\lambda+\delta\lambda^{(\lambda)}}{3!}\varphi^3_{0}(\vec{x'})<0|\chi_{I}(s,\vec{x})\chi_{I}(s',\vec{x'})|0>\nonumber\\+3\int ds'd^3x'\frac{\lambda+\delta\lambda^{(\lambda)}}{3!}\varphi_{0}(\vec{x'})<0|\chi_{I}(s',\vec{x'})\chi_{I}(s',\vec{x'})|0><0|\chi_{I}(s,\vec{x})\chi_{I}(s',\vec{x'})|0>\nonumber\\-\delta_{z}^{(\lambda)}\int ds'd^3x'\nabla^2\varphi_{0}(\vec{x'})<0|\chi_{I}(s,\vec{x})\chi_{I}(s',\vec{x'})|0>+\delta_{m}^{(\lambda)}\int ds'd^3x'\varphi_{0}(\vec{x'})<0|\chi_{I}(s,\vec{x})\chi_{I}(s',\vec{x'})|0>\nonumber\\
-\int ds'd^3x'\delta J_{0}^{(\lambda)}(\vec{x'})<0|\chi_{I}(s,\vec{x})\chi_{I}(s',\vec{x'})|0>=0~~~~~~~~~~~~
\eea
and we will obtain:
\bea
\delta J_{0}^{(\lambda)}(\vec{x'})=\frac{\lambda+\delta\lambda^{(\lambda)}}{3!}\varphi^3_{0}(\vec{x'})+\frac{\lambda+\delta\lambda^{(\lambda)}}{2}\varphi_{0}(\vec{x'})G(0)
-\delta_{z}^{(\lambda)}\nabla^2\varphi_{0}(\vec{x'})+\delta_{m}^{(\lambda)}\varphi_{0}(\vec{x'}).~~~~~~~~
\eea	
In continue, we write the equation (2.18) for this example:
\bea
\ddot{\phi}(t_{0},\vec{x})-\nabla^2\varphi_{0}(\vec{x})+m^2\varphi_{0}(\vec{x})+\frac{\lambda+\delta\lambda}{3!}<\psi_{0}|(\varphi_{0}(\vec{x})+\chi(\vec{x}))^3|\psi_{0}>\nonumber\\
+\delta_{z}\ddot{\phi}(t_{0},\vec{x})-\delta_{z}\nabla^2\varphi_{0}(\vec{x})+\delta_{m}\varphi_{0}(\vec{x})=0.
\eea
Since 
\be
<\psi_{0}|(\varphi_{0}(\vec{x})+\chi(\vec{x}))^3|\psi_{0}>=\varphi^3_{0}(\vec{x})+3\varphi_{0}(\vec{x})<\psi_{0}|\chi^2(\vec{x})|\psi_{0}>
+<\psi_{0}|\chi^3(\vec{x})|\psi_{0}>\nonumber\\
\ee
and its zeroth order is: $\varphi^3_{0}(\vec{x})+3\varphi_{0}(\vec{x})G(0)$, therefore the first order of the equation (3.11) is:
\bea
\ddot{\phi}(t_{0},\vec{x})-\nabla^2\varphi_{0}(\vec{x})+m^2\varphi_{0}(\vec{x})+\frac{\lambda+\delta\lambda^{(\lambda)}}{3!}(\varphi^3_{0}(\vec{x})+3\varphi_{0}(\vec{x})G(0))\nonumber\\
+\delta_{z}^{(\lambda)}\ddot{\phi}(t_{0},\vec{x})-\delta_{z}^{(\lambda)}\nabla^2\varphi_{0}(\vec{x})+\delta_{m}^{(\lambda)}\varphi_{0}(\vec{x})=0.
\eea
To keep the recent equation finite, we have to define renormalization conditions. It seems that, a completely classical field is a field with constant value in the direction of spacial coordinates. That means if a field has a maximum value in a spacial point, we are expecting the behaviors like particles from that fields in that point which is interpreted as quantum effects. Thus the renormalization condition is:\\ The equation of motion of the expectation value of the field is completely classical, at the time in which the expectation value of the field has a constant function in the direction of the spacial coordinates.\\ The recent condition determines the counter term coefficients in the equation (3.12) like this:  
\be
\delta\lambda^{(\lambda)}=\delta_{z}^{(\lambda)}=0~~ and ~~\delta_{m}^{(\lambda)}=-\frac{\lambda}{2}G(0)
\ee
and therefore the equation (3.10) changes to this:
\be
\delta J_{0}^{(\lambda)}(\vec{x})=\frac{\lambda}{3!}\varphi^3_{0}(\vec{x}).
\ee
And finally, the equation (3.11) for the first order and at the all moments, is:
\bea
\ddot{\phi}(t,\vec{x})-\nabla^2\phi(t,\vec{x})+m^2\phi(t,\vec{x})+\frac{\lambda}{3!}\phi^3(t,\vec{x})=0.
\eea
\section{discussion} 
In a quantum mechanic system, a ground state is an eigenstate of the Hamiltonian with the minimum eigenvalue. Quantum systems in normal mode, are in the ground states unless they are provoked, then they enter in the excited states. But the excited states in quantum mechanics have small life time and move back to the ground states. But in quantum field theory, we need excited states that make all the creatures of the world: The great bodies, the fields that fill the space, and their interactions between them. Such an excited state would be stable for long times, and maybe we call it, spontaneous excited state. This state has minimum energy while it is not the ground state. Its excitation is related to the sources has been turned on in the hamiltonian. But against the De Witt mechanism, they are no sources in the equation of the field operator. And for this reason, the external field apply in the classical equation of motion, without the need for turning off the sources.\\
 In this paper, our solutions have been obtained to the first order of interaction and our renormalization counter terms became as like as the one loop order of the ordinary renormalization conditions of $\frac{\lambda}{4!}\Phi^4$ action. In the next paper we want to compute the higher order of the equation of motion and also we want to explore the different renormalization conditions. It seems that different factors like boundary conditions, measurement  methods , classical self interactions and etc, are effective on the definition of the renormalization conditions.\\
  Here we have defined our constraints at the determined moment. In the future, we will try to check out that if any constraints could be defined not only at a moment but in a specified spacial spot, too. And also, is it possible if a hamiltonian density could be minimized to fix the behavior of the fields in a determined point of the space-time. We want to know if there is a local excitation in the field thoery.

\vspace*{1cm}

{\bf Acknowledgments}

I would like to thank dear..... for his useful discussions
and comments.

\end{document}